\documentclass[10pt,conference]{IEEEtran}
\IEEEoverridecommandlockouts

\usepackage{alltt}
\usepackage{eclbkbox}
\usepackage{cite}
\usepackage{amsmath,amssymb,amsfonts}
\usepackage{algorithmic}
\usepackage[pdftex]{graphicx,xcolor}
\usepackage{textcomp}
\def\BibTeX{{\rm B\kern-.05em{\sc i\kern-.025em b}\kern-.08em
    T\kern-.1667em\lower.7ex\hbox{E}\kern-.125emX}}

\renewcommand{\baselinestretch}{1.01}
\begin{document}

\title{Analyzing Code Comments to Boost Program Comprehension}

\author{\IEEEauthorblockN{1\textsuperscript{st} Yusuke Shinyama}
\IEEEauthorblockA{\textit{Department of Computer Science} \\
\textit{Tokyo Institute of Technology}\\
Tokyo, Japan \\
{\tt euske@sde.cs.titech.ac.jp}}
\and
\IEEEauthorblockN{2\textsuperscript{nd} Yoshitaka Arahori}
\IEEEauthorblockA{\textit{Department of Computer Science} \\
\textit{Tokyo Institute of Technology}\\
Tokyo, Japan \\
{\tt arahori@cs.titech.ac.jp}}
\and
\IEEEauthorblockN{3\textsuperscript{rd} Katsuhiko Gondow}
\IEEEauthorblockA{\textit{Department of Computer Science} \\
\textit{Tokyo Institute of Technology}\\
Tokyo, Japan \\
{\tt gondow@cs.titech.ac.jp}}
}

\maketitle

\begin{abstract}
We are trying to find source code comments that help programmers
understand a nontrivial part of source code. One of such examples
would be explaining to assign a zero as a way to ``clear'' a
buffer. Such comments are invaluable to programmers and identifying
them correctly would be of great help. Toward this goal, we developed
a method to discover explanatory code comments in a source code. We
first propose eleven distinct categories of code comments. We then
developed a decision-tree based classifier that can identify
explanatory comments with 60\% precision and 80\% recall. We analyzed
2,000 GitHub projects that are written in two languages: Java and
Python. This task is novel in that it focuses on a microscopic comment
(``local comment'') within a method or function, in contrast to the
prior efforts that focused on API- or method-level comments. We also
investigated how different category of comments is used in different
projects. Our key finding is that there are two dominant types of
comments: preconditional and postconditional. Our findings also
suggest that many English code comments have a certain grammatical
structure that are consistent across different projects.
\end{abstract}

\section{Backgrounds}

Source code comments are considered to be an indispensable part of
computer programs. Code comments are often used to make up for the
lack of proper software documentation. However, comments are also
considered as an elusive part of a computer program. Unlike the actual
code that is written in a programming language, code comments are
mostly written in natural language, which has a lot more freedom in
expressions and hence defies any sort of formal analysis or objective
testing.

We are particularly interested in source code comments that are inside
a function or method and explains how the code works at a microscopic
level. In this paper, we call them {\bf ``local comments''}.  Local
comments tend to describe things like ``the kind of data that is
stored in a certain variable'' or ``the assumption that is hold at a
certain point'' and so on. They normally describe a small part (often
just a few lines) of code and largely invisible from the official
documents, because they are too technical or obscure to most
users. Such comments, however, are often crucial for understanding a
tricky part of the code, as they give us a rare glimpse of its
developer's mind.  Programmers often need this kind of information for
diagnosing the software issues or extending its functions.

Note that this paper is a part of our bigger attempt that is to obtain
the semantic relationship between a source code and a program
function.  We plan to use the techniques presented in this paper to
analyze the function of an unknown program given its source
code. However, this future stage is not presented in this paper.

Our original goal was to leverage local comments for static analysis
of a program. Soon we discovered that finding good local comments
itself is a nontrivial task. One of such comments would be the
following:
\begin{center}
\begin{minipage}[b]{0.95 \linewidth}
\begin{alltt} \small
b.start = b.end = 0; // clear the ring buffer.
\end{alltt}
\end{minipage}
\end{center}
In this example, the actual operation of the code is just to assign
zeroes to two variables, \verb+b.start+ and \verb+b.end+.  But the
comment gives it a much richer meaning than just two assignments;
namely, {\it they are clearing the ring buffer}. In order to discover
this kind of examples, we need to know a couple of things. First, we
have to recognize that the above comment actually explains the code to
its left. Then we also have to recognize that the comment explains the
{\it function} of the code. To illustrate how this is not a trivial
problem, let us consider another example:
\begin{center}
\begin{minipage}[b]{0.9 \linewidth}
\begin{alltt}
// error occurred.
b.start = b.end = 0;
\end{alltt}
\end{minipage}
\end{center}
In this case, the role of the comment is entirely different. While
this comment might be still as important and relevant, it is no more
explaining what the code does. Rather, it is stating the reason why
this code needs to be executed. Obviously, there are several different
roles for code comments, but it was not clear how they are different
and what we can do to recognize them automatically.

There have been a few seminal attempts to categorize source code
comments. Padioleau et al. studied code comments in operating system
kernels and classified them by detailed topics \cite{padioleau:09}.
Steidl et al. proposed seven categories of comments \cite{steidl:13}.
They focused on macroscopic comments, such as a module header or
method description. Pascarella and Bacchelli further developed this
idea and proposed a hierarchical categories \cite{pascarella:17}. The
key difference between our work and these previous works is that we
primarily focus on a finer relationship between a code and its local
comment, with the aim of collecting the code/comment pairs for future
use.

Today, several industry guidelines \cite{javadoc, kramer:99} exist for
writing macroscopic, or non-local comments. However, the manner and
style of local comments are still pretty much up to a programmer's own
discretion. In {\it The Practice of Programming} \cite{kernighan:99},
Kernighan and Pike mentioned a couple of principles for writing code
comments such as ``Don't belabor the obvious'' or ``Clarify, don't
confuse'', but they did not go further.  As a result, analyzing local
comments is still believed to be hard and there have not been many
attempts in this area.

We initially set out collecting comments from popular GitHub projects
(repositories) and manually reviewing them. Throughout the process, we
recognized that there are some general unspoken rules for local
comments. Our own experiences as a programmer agree with this too;
each programmer develops their own ``grammar'' for comments, even when
there is no explicit instruction.  While the style of each local
comment still varies from project to project, we observed that there
is a certain tendency among them, which brings hope that we can
analyze them somewhat mechanically. This paper is our attempt to
propose a framework that allows a rigid analysis of local comments and
lay the groundwork for further research in this field.

\subsection{Contribution of Paper}

There are three contributions of this paper. Firstly, we propose
common structural elements of local comments written in natural
language. To our knowledge, this is the first attempt to automatically
analyze source code comments in this level of detail. Secondly, we
develop a machine learning algorithm that can identify each element of
a local comment with a reasonable accuracy. Thirdly, we show those
structures are roughly preserved across different projects and, to
some extent, in different languages.

More specifically, we try to answer the following Research Questions:
\begin{description}
\item [RQ1.] What are the common (syntactic or semantic) elements
  among source code comments written in natural language (English)?
\item [RQ2.] Can those elements be identified
  by a certain machine learning algorithm?
\item [RQ3.] How common those elements can be seen
  across different projects or programming languages?
\end{description}

The rest of this paper goes as follows: First, we present our model of
local comments that explain how a code works. Then we describe our
attempt to automatically identify them using a machine learning
algorithm (decision tree). And then we proceed to apply this method to
real projects and see if our model is relevant. We then discuss our
findings.  Finally, we briefly discuss the related work and state the
conclusion.

\section{Structure of Comments}
\label{sec:structure}

\begin{figure}
\begin{center}
\includegraphics[width=0.6\linewidth]{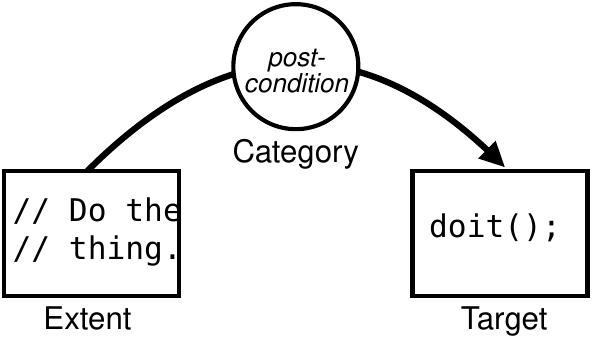}
\caption{Comment as Relationship}
\label{fig:relation}
\end{center}
\end{figure}

In this section, we present our model of source code comments in
attempting to answer our first Research Question.  In most cases,
there is a relationship between a comment and the code it
describes. Each relationship can be viewed as an arc that has three
elements (Fig. \ref{fig:relation}). The arc has its source (comment
itself), the destination (the target code), and a type of relationship
(comment category). In the rest of this section, we describe each
element one by one.

\subsection{Comment Extent}

The extent of a comment is a part of source code that can be
recognized as one ``chunk'' of explanatory text. This is less obvious
than it sounds because in many cases, a single explanatory text is not
necessarily expressed by a single comment tag. This is illustrated by
the following example:

\begin{center}
\begin{minipage}[b]{0.5 \linewidth}
\begin{alltt}
// This is still
// one sentence.
\end{alltt}
\end{minipage}
\end{center}

In many modern languages such as Java, C++ or Python, inline comments
are commonly used. Inline comments, or end-of-line comments, are a
type of comments that start with a comment tag (such as \verb+//+ or
\verb+#+) and continues until the end of the line. It is very common
that a single sentence is split into multiple inline comments because
programmers want to limit the length of each line to maintain its
readability.

In the rest of this paper, we define the extent of a comment as a
sequence of consecutive comment tags that can be taken as a single
continuous text. One could say a comment extent is a ``whole'' comment
rather than individual comment tags.  In theory, however, there could
be a disjoint comment set that forms one explanation. We have not
found such an example from the source codes we reviewed.

\subsection{Comment Target}

Comments are by nature aiming at a certain subject, as we often say
we comment {\it on something}. The same can be said for source code
comments.  However, there has not been a lot of discussion about what
are the actual targets of code comments or how they can be specified
in terms of a programming language syntax.

\begin{figure}
\begin{center}
\fbox{\includegraphics[width=0.9\linewidth]{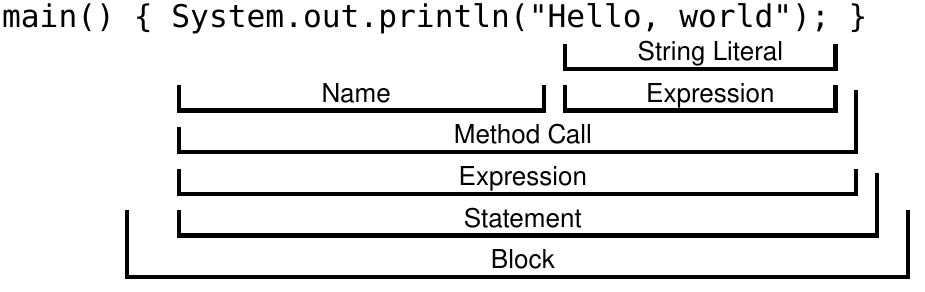}}
\caption{Parse Tree Example (Java)}
\label{fig:javaparse}
\end{center}
\end{figure}

In modern languages, a program source code is first transformed into a
parse tree. A typical parse tree consists of a number of syntax
elements that cover certain parts of the source code (Fig.
\ref{fig:javaparse}).  In many popular languages, however, comments
are treated as special tokens and not a part of a syntax tree. Java
Development Tools (JDT), a popular Java parser implementation, treats
source code comments as a special syntax element that belongs to the
entire file \cite{java-jdt}. In Python Abstract Syntax Tree (AST)
module, comments are discarded at the tokenization stage and
completely ignored by the subsequent parser. However, when a
programmer writes a comment, they often try to align it with the
existing language syntax.  Therefore it is natural to assume that
there is some way to express the target of a comment using some form
of formal syntax.

From manually reviewing 1,000 Java code comments, we have discovered
that most comment targets can be specified relative to its surrounding
syntax elements (Fig. \ref{fig:target}).  Every syntax element
within a parse tree has its start and end point, and comments are
sitting between two syntax elements. We found that comment targets can
be specified as one of the four types:

\begin{figure}
\begin{center}
\includegraphics[width=0.7\linewidth]{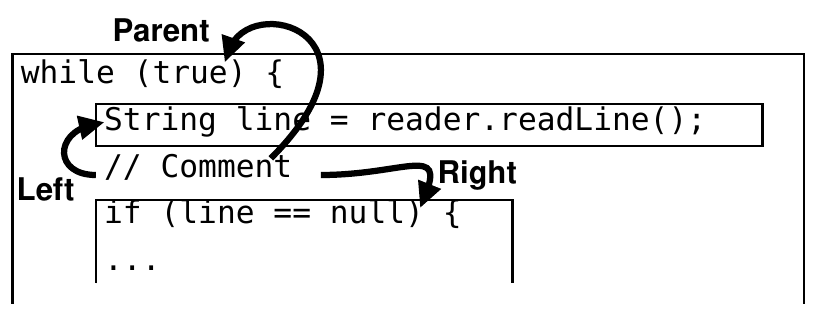}
\caption{Comment Target}
\label{fig:target}
\end{center}
\end{figure}

\begin{itemize}
\item Left : Comment is targeting the syntax element that ends
  immediately before the comment. When there are overlapping elements
  which ends at the same point, the element which has the longest
  span is chosen.  Note that this will change the size of a comment
  target depending on its position. For example, comment
{\small \begin{alltt}
  thread.join();  // Let the job finish.
\end{alltt}}
targets the entire statement that precedes it, whereas
{\begin{alltt} \scriptsize
  c.query(uri, DOWNLOAD, null /* selection */);
\end{alltt}}
only targets the expression (\verb+null+).
\item Right : Comment is targeting the syntax element that starts
  immediately after the comment. When there are overlapping elements
  which starts at the same point, the longest element is chosen.
  A typical example would look like this:
{\small \begin{alltt}
  // Copy the array.
  for (int i = 0; i < a.length; i++) \{
    b[i] = a[i];
  \}
\end{alltt}}
  where the comment targets the entire \verb+for+ block to its right.
\item Parent : Comment is targeting its parent element, i.e. the
  syntax element that contains the comment. This type of target is
  commonly seen in an \verb+if+ statement:

{\small \begin{alltt}
  if (obj == null) \{ // error
      return;
  \}
\end{alltt}}
  where the comment targets the entire then-block.
\item In-Place : Comment does not describe any code.
  The target is considered as the comment itself.
  Examples include metadata such as authors or copyright
  notices.

\end{itemize}

Note that not all comments have a target. A notable example is a
commented out code. Such a comment is considered to have {\bf
  ``In-Place''} target.  Out of 1,000 comments we have seen, only 3 of
did not fit in the above four criteria. In the rest of this paper, we
ignore these irregular targets.

\subsection{Comment Category}

A comment category represents the type of a relationship between a
comment and its target. While a comment extent and comment target are
mostly syntactic elements, a comment category involved some sort of
semantics. There are several existing works about comment categories
\cite{steidl:13, pascarella:17}, but we independently made our list of
categories that are suitable for local comments. The procedure of
making the category list is the following: We reviewed each comment
and checked if this comment can be in one of existing categories.  If
not, we regarded the comment as of a new category. After finishing
this process, eleven categories were formed. They are listed in
Tab. \ref{tab:categories}. Note that certain categories such as {\bf
  ``Guide''} or {\bf ``Meta Information''} are rarely used for local
comments, but left for the sake of completion.

\begin{table}
\caption{Comment Categories}
\begin{center}
\begin{tabular}{|p{0.95\linewidth}|}
  \hline
  {\bf Postcondition} \\
  \hspace{1ex} \begin{minipage}[b]{0.9\linewidth}
  {\small Conditions or effects that hold {\it after} the code is executed.
  Typically used for explaining ``what'' the code does.}
  {\scriptsize \begin{alltt}
    // create some test data
    Map<String, String> data = createTestData(testSize);

    // if we had a prior association,
    // restore and throw an exception
    if (previous != null) \{
        taskVertices.put(id, previous);
        ...
\end{alltt} }
  \end{minipage}
  \\ \hline
  {\bf Precondition} \\
  \hspace{1ex} \begin{minipage}[b]{0.9\linewidth}
  {\small Conditions that hold {\it before} the code is executed.
  This includes statements that hold {\it regardless} of the code execution.
  Typically used for explaining ``why'' the code is needed.}
  {\scriptsize \begin{alltt}
    // Unable to find the specidifed document.
    return Status.ERROR;

    if (myStatusBar != null) \{ //not welcome screen
      myStatusBar.addProgress(this, myInfo);
    \}
\end{alltt} }
  \end{minipage}
  \\ \hline
  {\bf Value Description} \\
  \hspace{1ex} \begin{minipage}[b]{0.9\linewidth}
  {\small Phrase that can be equated with a variable, constant or expression.}
  {\scriptsize \begin{alltt}
    addSourceFolders(
        SourceFolder.FACTORY,
        getSourceFoldersToInputsIndex(),
        false /* wantsPackagePrefix */,
        context);
\end{alltt} }
  \end{minipage}
  \\ \hline
  {\bf Instruction} \\
  \hspace{1ex} \begin{minipage}[b]{0.9\linewidth}
  Instruction for {\it code maintainers}. Often referred to as ``TODO'' comments.
  {\scriptsize \begin{alltt}
    // TODO Auto-generated catch block
    e.printStackTrace();
    Assert.fail("Failed");
\end{alltt} }
  \end{minipage}
  \\ \hline
  {\bf Guide} \\
  \hspace{1ex} \begin{minipage}[b]{0.9\linewidth}
  Guide for {\it code users}. Not to be confused with Instructions.
  {\scriptsize \begin{alltt}
    // Example: renderText(100, 100, FONT, 12, "Hello");
\end{alltt} }
  \end{minipage}
  \\ \hline
  {\bf Interface} \\
  \hspace{1ex} \begin{minipage}[b]{0.9\linewidth}
  Description of a function, type, class or interface.
  {\scriptsize \begin{alltt}
    // Comparison function
    class MyComparator implements Comparator \{
      public int compare(Object o1, Object o2) \{
      ...
\end{alltt} }
  \end{minipage}
  \\ \hline
  {\bf Meta Information} \\
  \hspace{1ex} \begin{minipage}[b]{0.9\linewidth}
  {\small Meta information such as author, date, or copyright.}
  {\scriptsize \begin{alltt}
    // from org.apache.curator.framework.
    // CuratorFrameworkFactory
    this.maxCloseWait = 1000;
\end{alltt} }
  \end{minipage}
  \\ \hline
  {\bf Comment Out} \\
  \hspace{1ex} \begin{minipage}[b]{0.9\linewidth}
  Commented out code. This type of comments does not have its target.
  {\scriptsize \begin{alltt}
    while ((m = ch.receive()) != null) \{
      //System.out.println(Strand.currentStrand());
      ...
\end{alltt} }
  \end{minipage}
  \\ \hline
  {\bf Directive} \\
  \hspace{1ex} \begin{minipage}[b]{0.9\linewidth}
  Compiler directive that isn't directed to human readers.
  {\scriptsize \begin{alltt}
    //CHECKSTYLE:OFF
    \} catch (final Exception ex) \{
    //CHECKSTYLE:ON
\end{alltt} }
  \end{minipage}
  \\ \hline
  {\bf Visual Cue} \\
  \hspace{1ex} \begin{minipage}[b]{0.9\linewidth}
  Text inserted just for the ease of reading.
  {\scriptsize \begin{alltt}
    //
    // Initialization key storage
    //
\end{alltt} }
  \end{minipage}
  \\ \hline
  {\bf Uncategorized} \\
  \hspace{1ex} \begin{minipage}[b]{0.9\linewidth}
  All other comments that don't fit the above categories.
  \end{minipage}
  \\ \hline
\end{tabular}
\label{tab:categories}
\end{center}
\end{table}

\subsubsection{Manual Annotation Experiment}
\label{subsubsec:manual}

\begin{figure}
\begin{center}
\fbox{\includegraphics[width=0.7\linewidth]{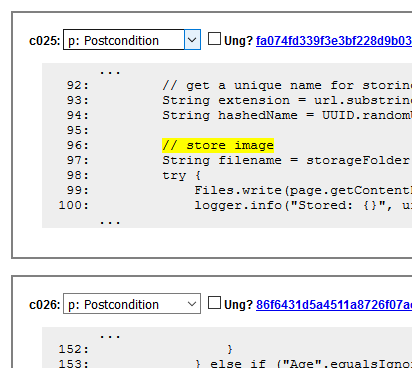}}
\caption{Screenshot of Annotation Tool}
\label{fig:screenshot}
\end{center}
\end{figure}

In order to verify the relevance of our definition of the comment
categories, we have conducted a manual annotation experiment.  The
three authors of this paper participated this experiment.  Each
participant is given a list of guidelines shown in Tab.
\ref{tab:categories} and instructed to choose a category for 100 Java
code snippets. The annotation tool is implemented as a form of Web
application where a participant can choose the categories from a
drop-down menu (Fig. \ref{fig:screenshot}). The time taken for each
choice is recorded. The categories obtained here are later used as a
test set for measuring the performance of our classifiers.

The code snippets given in this experiment are randomly chosen from
popular 100 GitHub projects. All the participants receive the same set
of snippets. Each snippet contains a comment and its neighboring
lines (four lines before and after the comment). Since some questions
might require a participant to study the source code in-depth, a link
is provided for each snippet, which leads to the original GitHub
repository where the participant can view a wider range of the source
code, or other project files if necessary.

There are three participants in this experiment. One of them is a
doctoral student and the other two are faculty members. They are all
male, and their age ranges from 37 to 51. The total time spent for
this experiment ranges from 30 minutes to 2 hours per person.  Their
median time for each question ranges from 10 seconds to 30 seconds.

\subsubsection*{Measuring the Agreement Ratio}

We used Fleiss' Kappa as the ratio of inter-rater agreement.  Fleiss'
Kappa is commonly used for measuring agreement between $N$ people
where $N \ge 3$. In case of $N = 2$, Cohen's Kappa is typically used. Both
methods are based on the same principle as we illustrate in the
following paragraphs.

When there is no gold standard for answers, the only ratio we can
measure for agreement is how many times people choose the same
category.  However, people choosing the same category on some question
does not guarantee that they always agree on every question. The idea
behind Cohen's Kappa is to prevent one type of answers from
accidentally dominating the entire agreement ratio. Technically, this
is done by discounting the agreement for categories that are
frequently chosen.

Fleiss' Kappa is an extention to Cohen's Kappa for three or more
people.  This is calculated as follows: Assume that there is a
complete graph that connects all the participants (which is a
triangle, in our case), and count the number of edges where both
participants on the edge agree on the answer. Then discount the
agreement on a frequent category in the same manner of Cohen's Kappa.
We calculated the Fleiss' Kappa for our expriment as $K=0.491$. By a
commonly used guideline, this is considered as ``moderate
agreement''. The confusion matrix is shown in Tab.
\ref{tab:agreements}. The agreement ratio on comment extents or
comment targets were not measured.

\begin{table}
\caption{Agreements on Manual Annotations}
\begin{center}
{\small
\begin{tabular}{|l|r|r|r|r|r|r|r|r|r|}
  \hline
  Category      & Po & Pr & Co & Vi & Va & In & Di & Gu & Un \\
  \hline
  Postcondition & {\bf 75} &    &    &    &    &    &    &    &    \\
  Precondition  & 37 & {\bf 37} &    &    &    &    &    &    &    \\
  Comment Out   &  0 &  0 & {\bf 12} &    &    &    &    &    &    \\
  Visual Cue    & 17 & 11 &  0 & {\bf 11} &    &    &    &    &    \\
  Value Descr.  & 10 &  7 &  0 &  1 & {\bf 15} &    &    &    &    \\
  Instruction   &  2 &  4 &  0 &  1 &  0 & {\bf 14} &    &    &    \\
  Directive     &  0 &  0 &  0 &  5 &  0 &  0 &  {\bf 8} &    &    \\
  Guide         &  0 &  0 &  0 &  1 &  0 &  0 &  1 &  {\bf 0} &    \\
  Uncategorized &  6 &  5 &  0 & 10 &  0 &  1 &  1 &  0 &  {\bf 9} \\
  \hline
\end{tabular}
}
\label{tab:agreements}
\end{center}
\end{table}

We have found that there is a relatively high chance of disagreement
between {\bf ``Precondition''} and {\bf ``Postcondition''} categories.
We think there are three major reasons for this: Firstly, both
categories are inherently tricky because they require a deep
understanding of the code and comment. Sometimes the participants had
to view a much wider range of the code to get the proper context for a
snippet.  Secondly, some comments are long enough that can actually
have multiple purposes. Although the participants are asked to choose
the most dominant category, it is sometimes not clear which category
fits the best. And thirdly, the code quality and style vary between
different projects.

\section{Building Classifiers}
\label{sec:classifiers}

As the answer to our second Research Question, we now describe our
attempt to build machine learning classifiers that identify the three
elements described in the previous section. We use C4.5 decision tree
algorithm \cite{quinlan:93}. A decision tree algorithm is efficient
and easy to implement, but the feature set we use is relatively
complex, as explained later.  We particularly like its property that
the obtained tree is to some extent human readable, allowing us to
investigate which feature works the best.

\begin{figure*}
\begin{center}
\fbox{\includegraphics[width=0.7\linewidth]{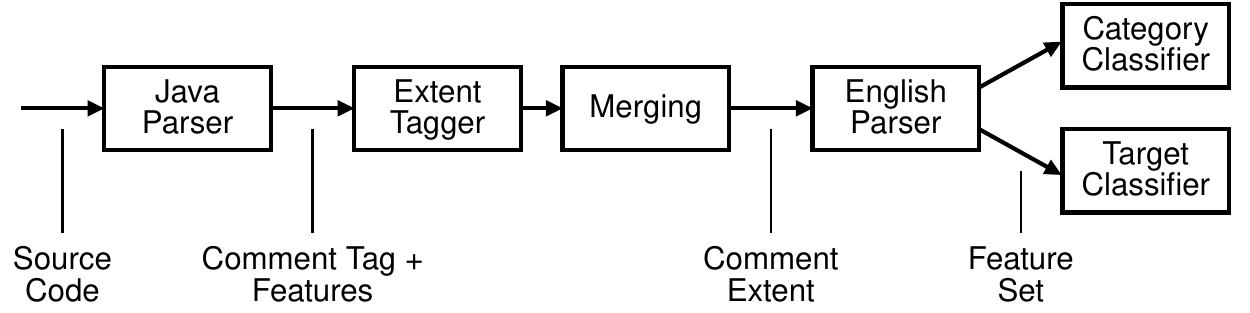}}
\caption{System Pipeline}
\label{fig:pipeline}
\end{center}
\end{figure*}

We built three different classifiers for each element: Extent, Target
and Category. Fig. \ref{fig:pipeline} shows the overall system
architecture. An input source code is first processed by a Java
parser.  We use Java Development Tools (JDT) \cite{java-jdt}
here. Then the parse tree is fed into the Extent Tagger. This is the
first classifier that identifies the beginning and end of a comment
extent. Extent Tagger needs to be applied prior to the other two
classifiers (Target and Category) because the later classifiers
require an entire comment extent rather than individual
comments. After this stage, inline comments are grouped into one chunk
and each comment extent has one continuous text. Then a natural
language parser is applied. We use an English parser with an
assumption that the majority of source code comments are written in
English\footnote{Out of 1,000 comments we have reviewed, we found 38
  of them were actually written in Chinese. All other comments were in
  English.}.  Finally, the combined features are fed into Target and
Category classifiers respectively.  In the following subsections, we
describe each classifier in the pipeline.

\subsection{Recognizing Comment Extent}

\begin{figure}
\begin{center}
\fbox{\includegraphics[width=0.9\linewidth]{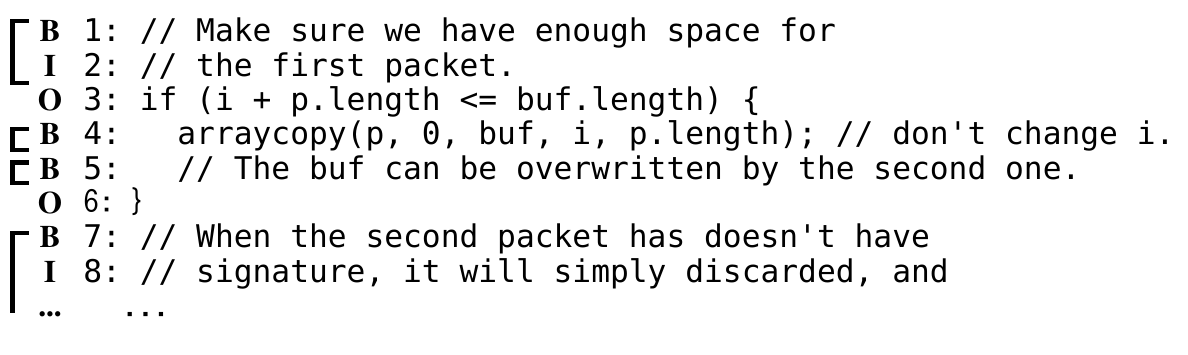}}
\caption{IOB Notation Example}
\label{fig:biochunk}
\end{center}
\end{figure}

Extent Tagger is a decision tree-based classifier that marks the
beginning and end of each comment extent. In order to perform this
as a classification task, we use IOB notation \cite{slp}
\footnote{It is sometimes called ``BIO notation''.}.  Each comment tag
is assigned with one of the three tags: ``I'' (Middle of an extent),
``O'' (Outside of an extent) or ``B'' (Beginning of an extent).  A
tagging example is shown in Fig. \ref{fig:biochunk}.  Note that the
comments at line 4. and 5. are regarded as separate even though they
are consecutive, hence giving a ``B'' tag to both lines.  Since we
only deal with comment tags, only ``B'' and ``I'' tags are considered
in an actual process. IOB notation is a common technique to mark the
boundary of an object sequence, and it is used in many natural
language processing tasks.

The actual mechanism of Extent Tagger works as follows: first, a
preprocessing is applied to each comment tag and its surrounding
syntax elements are identified. They are listed as a set of discrete
features. Since comments are not well integrated in a parse tree, we
process them separately and combine them based on its text location
within the source file. To our knowledge, this is the first attempt to
use precise relationship between a language parse tree with its
comments for categorizing comments.  Then the distance between each
comment and its neighboring comments is calculated. They are expressed
as a number of lines (vertical distance) and a number of columns
(horizontal distance). Finally, these features are fed into the
classifier and the IOB tag is identified.  The list of features used
is shown in Tab. \ref{tab:features-extent}.

\begin{table}
\caption{Classifier Features (Extent)}
\begin{center}
\begin{tabular}{|l|p{0.7\linewidth}|}
\hline
Feature & Description \\
\hline
\verb+DeltaRows+    & Distance in lines from a previous comment. \\
\verb+DeltaCols+    & Difference in columns from a previous comment. \\
\verb+DeltaLeft+    & Difference in columns between a comment and syntax element. \\
\verb+LeftSyntax+   & Syntax element left to the comment. \\
\verb+RightSyntax+  & Syntax element right to the comment. \\
\verb+ParentSyntax+ & Parent syntax element of the comment. \\
\hline
\end{tabular}
\label{tab:features-extent}
\end{center}
\end{table}

\subsection{Identifying Comment Target and Category}

After the Extent Tagger stage, comments are merged into one continuous
text. Then Part-of-speech (POS) tagging is applied. The POS tagger
assigns one of 36 POS tags to each word, such as \verb+VBZ+ (verb, 3rd
person) or \verb+NNS+ (plural noun) \cite{ptb}.  We use CoreNLP
natural language processing toolkit \cite{corenlp}. At this point, all
the words in a comment extent is assigned with a POS tag and they can
be used as features. We add a few extra binary features
(``HasSymbol'') using a regular expression pattern to detect if the
text includes a symbol that is commonly used in a program code. These
extra features are manually crafted and expected to help the
classifier to identify characteristics of certain categories.  Then we
independently apply two classifiers to this feature set and identify
the target and category of each comment extent. The list of used
features is shown in Tab. \ref{tab:features-category}.

\begin{table}
\caption{Classifier Features (Category)}
\begin{center}
\begin{tabular}{|l|p{0.7\linewidth}|}
\hline
Feature & Description \\
\hline
\verb+LeftSyntax+   & Syntax element left to the comment. \\
\verb+RightSyntax+  & Syntax element right to the comment. \\
\verb+ParentSyntax+ & Parent syntax element of the comment. \\
\verb+HasSymbol+    & Does the comment text include a symbol? \\
\verb+PosTagFirst+  & POS tag of the first word of the comment. \\
\verb+PosTagAny+    & Does the comment text include a certain POS tag? \\
\verb+WordFirst+    & First word of the comment text.  \\
\verb+WordAny+      & Does the comment text include a certain word? \\
\hline
\end{tabular}
\label{tab:features-category}
\end{center}
\end{table}

\begin{figure}
\begin{center}
\fbox{\includegraphics[width=0.6\linewidth]{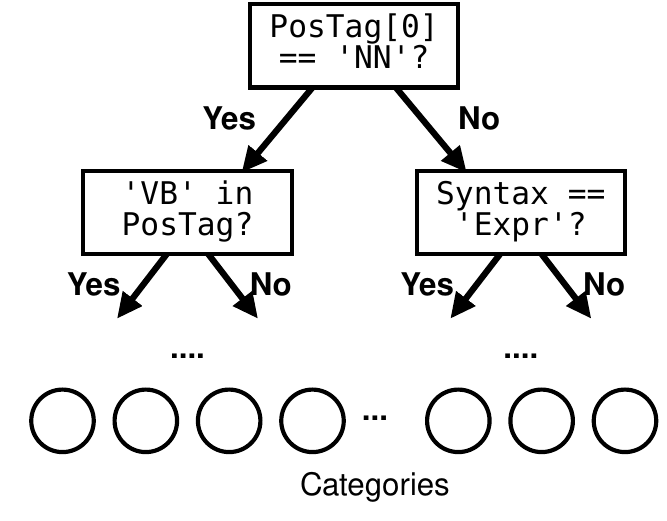}}
\caption{Decision Tree Example}
\label{fig:decisiontree}
\end{center}
\end{figure}

Our C4.5 implementation is fairly straightforward. The way that the
decision tree learner works is following: it scans all the input
examples and searches a feature that split the given examples the
best. In C4.5 algorithm, this means that a split with the maximum
information gain is chosen. The algorithm starts with the most
significant feature, and then repeatedly splits the subtrees until it
meets a certain predefined cutoff criteria; an important feature tends
to appear at the top of the tree, and as it descends to its nodes a
less significant feature appears. In general, setting the cutoff
threshold too small causes a tree overfitting problem, while setting
it too large makes it underfitting. In our experiment, we found that
setting the minimum threshold to 10 examples produced the best
results. The more detailed mechanism is described in
\cite{quinlan:93}. Fig. \ref{fig:decisiontree} shows a sample
decision tree. Once the decision tree is constructed, it is converted
to simple if-then clauses, so that the actual classification can
be performed efficiently.

\section{Experiments}

In this section, we describe our experimental setup and its
results. We have measured the performance of the three classifiers we
described above. We first describe our data set and then present the
experimental result with Java source codes. Then, to measure the
generality of our model, we apply the same classifier to another
language, Python. Finally, we apply our method to a wider range of
projects and show its findings.

\subsection{Data Set}
\label{subsec:dataset}

As a data set for the experiments, we selected the top 1,000 GitHub
projects by popularity (the number of Stars)\footnote{The data set
  was retrieved in July, 2017 (Java) and November, 2017 (Python)
  respectively.}. The overall size of the data set is listed in Tab.
\ref{tab:expdata}.  We then parsed all the Java files in each project
and randomly chose 1,000 comments\footnote{We enumerated
  all the comments of the above projects, shuffle them,
  then pick the first 1,000 comments while limiting
  the maximum number of comments per source file to three.
  This way, an unusually large source code does not affect
  the overall distribution, while large projects with many
  source code files can still be more representing than
  smaller projects.} as a training set for the
classifiers. These 1,000 comments were manually annotated for the
three elements described in Section \ref{sec:structure}.  The
frequency of each comment target and category in the training set is
listed in Tab. \ref{tab:trainset}. We then chose another 100 comments
independently and used them for the manual tagging experiment
described in Section \ref{subsubsec:manual}. The result of the tagging
experiment was further narrowed as some comment had no agreement in
category (all the participants chose different categories). In the
end, the remaining 84 comments were used as a test set. The distribution
of categories in the training set and test set is similar. The
Kullback-Leibler distance $D_{KL}$ between two distribution is 0.11. This was
calculated as
\[ \sum P_i \log \frac{P_i}{Q_i} \]
where $P_i$ and $Q_i$ are the probability of each category in
the training set and test set, respectively.

\begin{table}
\caption{Summary of Experimental Data}
\begin{center}
\begin{tabular}{|l|r|r|r|r|}
  \hline
  Language & Projects & Files & SLOC & Comments \\
  \hline
  Java & 1,000 & 480,600 & 63,224,880 & 4,049,628 \\
  Python & 990 & 160,844 & 29,070,278 & 2,215,683 \\
  \hline
\end{tabular}
\label{tab:expdata}
\end{center}
\end{table}

\begin{table}
\caption{Targets and Categories in Training Set}
\begin{center}
\begin{minipage}[b]{0.4\linewidth}
\begin{tabular}{|l|r|}
  \hline
  Target & Freq. \\
  \hline
  Right & 706 \\
  Left & 92 \\
  Parent & 74 \\
  In-Place & 125 \\
  Others & 3 \\
  \hline
  Total & 1,000 \\
  \hline
\end{tabular}
\end{minipage}
\begin{minipage}[b]{0.4\linewidth}
\begin{tabular}{|l|r|}
  \hline
  Category & Freq. \\
  \hline
  Postcondition & 610 \\
  Precondition & 148 \\
  Value Description & 67 \\
  Comment Out & 56 \\
  Instruction & 42 \\
  Visual Cue & 38 \\
  Directive & 26 \\
  Metadata & 5 \\
  Uncategorized & 8 \\
  \hline
  Total & 1,000 \\
  \hline
\end{tabular}
\end{minipage}
\label{tab:trainset}
\end{center}
\end{table}

\subsection{Experimental Results}

We now present the performance of our classifiers.  The first
classifier, Extent Tagger, had 97.7\% accuracy per comment tag.  The
Target classifier had 70\% accuracy per comment extent.  For the
Category classifier, we measured its performance for each category,
which is listed in Tab. \ref{tab:result_java}. Although the accuracy
of the classifier is varying depending on its category, it has a
reasonable performance (61\% precision and 89\% recall) for
the {\bf ``Postcondition''} category, which was our original purpose
for this research. Since there was no comment that was classified
as the {\bf ``Metadata''} category, its column is left out from both tables.

\begin{table}
\caption{Classifier Performance (Category, Java)}
\begin{center}
%
\begin{tabular}{|l|r|r|r|r|r|r|r|r|}
  \hline
  Category      & Po & Pr & Co & Vi & Va & In & Di & Cls. \\
  \hline
  Postcondition & {\bf 31} &  3 &  1 &  0 &  0 &  0 &  0 & 35\\
  Precondition  &  8 & {\bf 10} &  0 &  0 &  1 &  0 &  0 & 19 \\
  Comment Out   &  0 &  0 & {\bf 3}  &  0 &  1 &  0 &  0 & 4 \\
  Visual Cue    &  3 &  0 &  0 & {\bf 6}  &  0 &  0 &  0 & 9 \\
  Value Descr.  &  4 &  1 &  0 &  0 & {\bf 2}  &  0 &  0 & 7 \\
  Instruction   &  3 &  1 &  1 &  1 &  0 & {\bf 0}  &  0 & 6 \\
  Directive     &  2 &  0 &  0 &  1 &  0 &   0 & {\bf 1} & 4 \\
  \hline
  Answer        & 51 & 15 &  5 &  8 &  4 &  1 &  0 & 84 \\
  \hline
\end{tabular}

\vspace{1em}

\begin{tabular}{|l|r|r|r|}
  \hline
  Category & Precision & Recall & F1 \\
  \hline
  Postcondition & 0.61 (31/51) & 0.89 (31/35) & 0.72 \\
  Precondition  & 0.67 (10/15) & 0.53 (10/19) & 0.59 \\
  Comment Out   & 0.60 (3/5) & 0.75 (3/4) & 0.67 \\
  Visual Cue    & 0.75 (6/8) & 0.67 (6/9) & 0.71 \\
  Value Descr.  & 0.50 (2/4) & 0.29 (2/7) & 0.36 \\
  Instruction   & 0.00 (0/0) & 0.00 (0/6) & 0.00 \\
  Directive     & 1.00 (1/1) & 0.25 (1/4) & 0.40 \\
  \hline
\end{tabular}
\label{tab:result_java}
\end{center}
\end{table}

\subsection{Adapting to Another Language}
\label{subsec:python}

After experimenting with Java source codes, we applied the obtained
classifier to another language, Python.  This is done by applying a
rather straightforward transformation to the features. More
specifically, we converted the name of Java syntax elements in
features like \verb+LeftSyntax+, \verb+RightSyntax+ and
\verb+ParentSyntax+ into its Python counterparts by simply replacing
them (Tab. \ref{tab:transform_java}). All other features in the
decision tree were kept intact. Note that this decision tree was
originally obtained for Java source codes, so we did not make any
training data for this experiment. We manually annotated 100 comments
in Python for the test set in the same manner described in Section
\ref{subsec:dataset}.  We use the Python Abstract Syntax Tree (AST)
module for the parser \cite{python-ast}.

\begin{table}
\caption{Feature Transformation from Java to Python}
\begin{center}
\begin{tabular}{|l|l|}
  \hline
  Java Syntax & Python Syntax \\
  \hline
  \verb+SimpleName+ & \verb+Name+ \\
  \verb+MethodDeclaration+ & \verb+FunctionDef+ \\
  \verb+ExpressionStatement+ & \verb+Expr+ \\
  \verb+IfStatement+ & \verb+If+ \\
  \verb+MethodInvocation+ & \verb+Call+ \\
  \verb+ForStatement+ & \verb+For+ \\
  \verb+StringLiteral+ & \verb+Str+ \\
  \verb+NumberLiteral+ & \verb+Num+ \\
  \verb+ArrayInitializer+ & \verb+Tuple+ \\
  \hline
\end{tabular}
\label{tab:transform_java}
\end{center}
\end{table}

Tab. \ref{tab:result_python} shows the results.  Note that despite
the overall degradation of the performance in all categories, the
accuracy for {\bf ``Postcondition''} category stayed at a reasonable
level.  This result suggests that this type of comments have the same
characteristic in both languages, and thus implies the existence of
the universal ``grammar'' for them.

\begin{table}
\caption{Classifier Performance (Category, Python)}
\begin{center}
%
\begin{tabular}{|l|r|r|r|r|r|r|r|r|}
  \hline
  Category      & Po & Pr & Co & Vi & Va & In & Di & Cls. \\
  \hline
  Postcondition & {\bf 35} & 10 &  5 &  1 &  0 &  0 &  2 & 53\\
  Precondition  & 14 & {\bf 8}  &  3 &  2 &  1 &  2 &  1 & 31 \\
  Comment Out   &  0 &  0 & {\bf 1}  &  0 &  0 &  0 &  0 & 1 \\
  Visual Cue    &  3 &  0 &  0 & {\bf 0}  &  0 &  0 &  0 & 3 \\
  Value Descr.  &  3 &  3 &  0 &  2 & {\bf 1}  &  0 &  0 & 9 \\
  Instruction   &  1 &  0 &  1 &  0 &  0 & {\bf 1}  &  0 & 3 \\
  Directive     &  0 &  0 &  0 &  0 &  0 &   0 & {\bf 0} & 0 \\
  \hline
  Answer        & 56 & 21 & 10 &  5 &  2 &  2 &  4 & 100 \\
  \hline
\end{tabular}

\vspace{1em}

\begin{tabular}{|l|r|r|r|}
  \hline
  Category & Precision & Recall & F1 \\
  \hline
  Postcondition & 0.63 (35/56) & 0.66 (35/53) & 0.64 \\
  Precondition  & 0.38 (8/21) & 0.26 (8/31) & 0.20 \\
  Comment Out   & 0.10 (1/10) & 1.00 (1/1) & 0.18 \\
  Visual Cue    & 0.00 (0/5) & 0.00 (0/3) & 0.00 \\
  Value Descr.  & 0.50 (1/2) & 0.11 (1/9) & 0.11 \\
  Instruction   & 0.25 (1/4) & 0.33 (1/3) & 0.17 \\
  Directive     & 0.00 (0/2) & 0.00 (0/0) & 0.00 \\
  \hline
\end{tabular}
\label{tab:result_python}
\end{center}
\end{table}

\subsection{Applying to Large Corpora}

\begin{figure}
\begin{center}
\fbox{\includegraphics[width=0.9\linewidth]{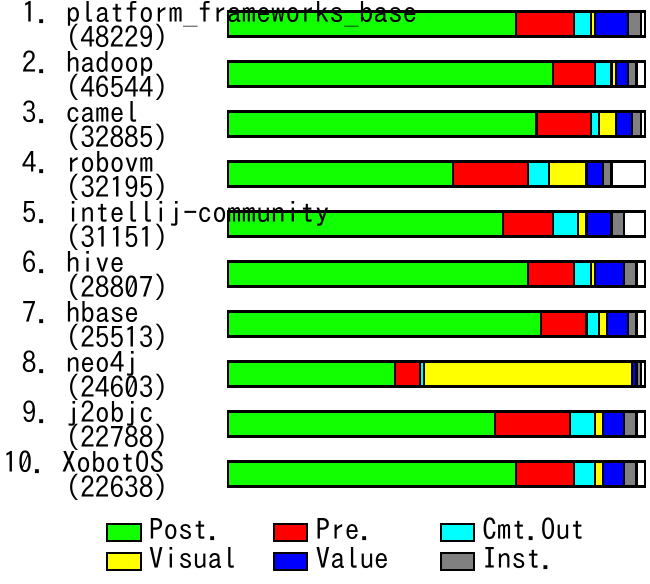}}
\caption{Comment Categories (Java)}
\label{fig:cat_java}
\end{center}
\end{figure}

\begin{figure}
\begin{center}
\fbox{\includegraphics[width=0.9\linewidth]{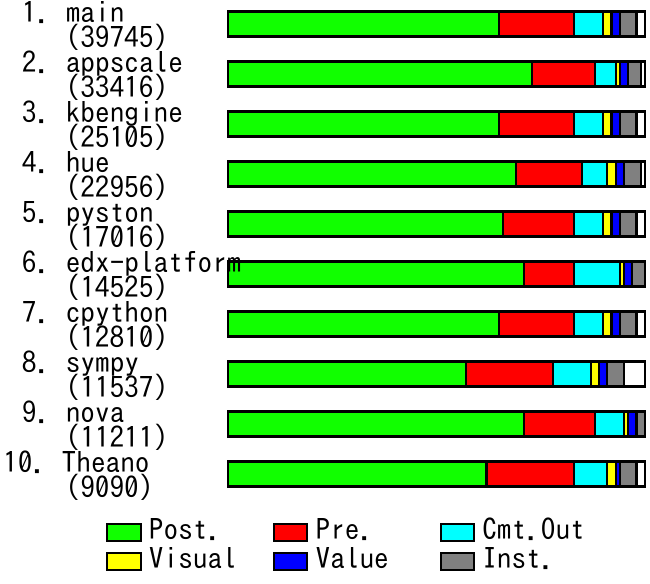}}
\caption{Comment Categories (Python)}
\label{fig:cat_python}
\end{center}
\end{figure}

We applied our classifier to a number of GitHub repositories to see
the overall tendency of comment categories in various projects, both
in Java and in Python. Fig. \ref{fig:cat_java} and
\ref{fig:cat_python} show the categories of the top 10 projects by the
number of comments (shown in the parentheses). We can see most
projects have a comparable ratio of categories. This answers our third
Research Question.

A notable exception in this figure is \verb+neo4j+ project, which has
an unusually high ratio of {\bf ``Visual Cue''} categories. It turned
out that this project has a high volume of unit testing codes (about
the 30\% out of the 900k lines of code).  According to the unit test
convention in Java, each unit test should have sections marked by
comments such as \verb+Given+, \verb+When+, and \verb+Then+. These
comments were eventually recognized as a visual cue by our classifier.

In an attempt to further explore our results, we extracted the most
commonly used ``verb + noun'' pairs used in the {\bf ``Postcondition''}
comments.  The results are shown in Tab. \ref{tab:words_java} and
\ref{tab:words_python}. It turned out that the most common phrase in
Java comments across all the projects is ``\verb+do nothing+''.  We also
tried to extract a few anecdotal code snippets which have ``\verb+clear+''
and ``\verb+buffer+'' in its corresponding comments. The obtained snippets
are shown in Fig. \ref{fig:clear_buffer}. This demonstrates one of
ways to utilize the proposed method for our original goal; by focusing
on postconditional comments, we can discover a number of nontrivial
ways where a buffer is ``cleared''.

\begin{table}
\caption{Top Verb+Noun Pairs (Java)}
\begin{center}
\begin{tabular}{|l|r|}
\hline
Verb + Noun & Projects \\
\hline
\verb+do nothing+ & 332 \\
\verb+throw exception+ & 170 \\
\verb+set default+ & 161 \\
\verb+add list+ & 154 \\
\verb+do anything+ & 146 \\
\verb+set value+ & 140 \\
\verb+use default+ & 122 \\
\verb+have value+ & 119 \\
\verb+create file+ & 119 \\
\verb+create list+ & 116 \\
\hline
\end{tabular}
\label{tab:words_java}
\end{center}
\end{table}

\begin{table}
\caption{Top Verb+Noun Pairs (Python)}
\begin{center}
\begin{tabular}{|l|r|}
\hline
Verb + Noun & Projects \\
\hline
\verb+create object+ & 149 \\
\verb+get list+ & 143 \\
\verb+get data+ & 134 \\
\verb+do anything+ & 133 \\
\verb+do nothing+ & 130 \\
\verb+get name+ & 129 \\
\verb+keep track+ & 128 \\
\verb+raise exception+ & 128 \\
\verb+write file+ & 123 \\
\verb+create file+ & 123 \\
\hline
\end{tabular}
\label{tab:words_python}
\end{center}
\end{table}

\begin{figure}
\begin{center}
\begin{breakbox}
\begin{alltt}
{\bf{OpenGrok/.../UtilTest.java}}
{\scriptsize  out.getBuffer().setLength(0); // clear buffer}

{\bf{atlas/.../BaseLayer.java}}
{\scriptsize  // Clear the off screen buffer. This is
  // necessary for some phones.
  canvas.drawRect(0, 0, canvas.getWidth(),
                  canvas.getHeight(), clearPaint);}

{\bf{druid/.../LimitedBufferGrouper.java}}
{\scriptsize  // clear the used bits of the first buffer
  for (int i = 0; i < maxBuckets; i++) \{
      subHashTableBuffers[0].put(
          i * bucketSizeWithHash, (byte)0);
  \}}

{\bf{hadoop/.../Shell.java}}
{\scriptsize  // clear the input stream buffer
  String line = inReader.readLine();
  while(line != null) \{
    line = inReader.readLine();
  \}}
\end{alltt}
\end{breakbox}
\caption{``Clear Buffer'' Examples (Java)}
\label{fig:clear_buffer}
\end{center}
\end{figure}

\section{Discussions}

In this section, we briefly discuss our experimental results in the
previous section. First of all, the performance of our Category
classifiers is worse than expected. This is probably due to the
similar reasons why the agreement ratio of manual annotations was bad;
the distinction between ``Precondition'' and ``Postcondition'' is
inherently tricky, and the comments were in varying qualities.

One of the ways to improve the performance is to use more
features. From an inspection of the obtained decision tree for the
classifier, we have discovered that \verb+POSTagFirst+ is the most
significant feature used for identifying a category, and programming
language syntax matters less. This is probably why it worked for both
Java and Python.  This roughly corresponds to our experience, as many
``Postcondition'' comments start with an imperative verb, as in the
form of ``do this''. Therefore, we can expect some performance gain by
using more natural language based features. For example, we could use
a full English parse tree as a feature, which would allow the
classifier to recognize a more complex phrase in a consistent manner.

Throughout the experiments, we have observed that the ratio of each
comment category is mostly unchanged across different projects and
different languages (Java and Python). The ratio of the
``Postcondition'' comments usually ranges from 60\% to 70\%, and the
ratio of the ``Precondition'' comments ranges from 10\% to 20\%.
Although they are not precisely comparable, we think that our
``Postcondition'' and ``Precondition'' categories roughly correspond
to the ``Summary'' and ``Expand'' categories described by Pascarella
and Bacchelli \cite{pascarella:17}, which reported a similar ratio
(about 7:2) across six Java projects. Assuming the ratio of each
category is not significantly different from projects to projects, our
result means that our classifier can consistently find the same
categories in both Java and Python projects, suggesting the existence
of universal ``grammar'' for source code comments.

\section{Threats to Validity}

There are a couple of threats to the validity of our conclusion. As
for the threats to the internal validity, we are aware that our
definition of comment targets or categories might still be
incomplete. The validity of our eleven comment categories can be
measured with the number of uncategorized comments (21 out of 300
annotations) as well as the inter-rater agreement ratio. One could
argue that the number of reviewed examples or the degree of agreement
among the manual annotations is weak. This is especially true for the
distinction between ``Precondition'' and ``Postcondition'' categories
(as stated in Section \ref{subsubsec:manual}). The small number of
annotators could also be a problem. The annotators might be
biased. However, the observed Fleiss' Kappa was $K=0.491$ (moderate
agreement), which is not bad.  As for the design of classifiers, we
have assumed that a comment target and comment category is independent
to each other (Section \ref{sec:classifiers}); this might not be the
case. Also, we treated that each comment extent as independent from
other comments, i.e. the interpretation of one comment extent is not
affected by other comments in the source code. Realistically, this is
very unlikely. We often see that a comment uses names, ideas, or terms
introduced by other comments in the same file or other external
files. This could potentially affect the classification results.

As for the threats to the external validity, one could argue that the
amount of test data is not enough. Indeed, the test set for Java
comment categories had only 84 cases (Section
\ref{subsec:dataset}). However, the distribution of categories from
the training set and test set was not very different, as shown in
Section \ref{subsec:dataset}. Another threat is that both training set
and test set are highly skewed in their categories. For now, our main
focus is to distinguish the two most major categories
(``Precondition'' and ``Postcondition''), so the predictions for
smaller categories should be treated carefully. We plan to address
this issue further in future.  The test set for Python comments was
equally small. Also, one could argue that the application of our
classifier to a Python source code did not work well, citing its lower
score (Section \ref{subsec:python}). In fact, Python language supports
\verb+docstring+, another mechanism for software documentation.
However \verb+docstring+ texts are usually only applied for a class or
method level description, but not for local comments. So we do not
think it affects the results.

Another threat to the external validity is the way we used GitHub
projects. They are all open sourced software, which could bias
the use of code comments. It is also known that there is a high
variation in its code quality in GitHub projects. A carefully
selected set of projects might give a different outcome.

\section{Related Work}

Source code comments have been an active target of research, but there
is no prior work that explores the semantics of local comments for the
purpose of obtaining nontrivial comments in an empirical manner.
While there are a group of people who advocate ``self-documenting
code'' \cite{selfdocumenting}, code comments are still considered as
an irreplaceable way to express programmers' intention
\cite{raskin:05}.  Padioleau et al. studied source code comments from
operating systems and concluded that programmers use comments when
they cannot express their intention in any other way
\cite{padioleau:09}. They also classified the comments by detailed
topics, such as ``error code'' or ``lock related''.  Comment
classification has been actively studied by Pascarella and Bacchelli
\cite{pascarella:17}. In terms of writing comments, Kramer provided a
case study of Java programmers who wrote Javadoc comments for Java API
at Sun Microsystems \cite{kramer:99}.

There are numerous attempts to mine source code comments to get extra
intelligence about a program. Jiang and Hassan examined the effects of
stale comments and bugs \cite{jiang:06}. Tan et al. presented a clever
approach that automatically finds lock-related bugs by obtaining
special patterns from code comments \cite{tan:07}. Ying et
al. explored ``TODO'' comments, or task comments, as a way of
programmers' communication to their coworkers \cite{ying:05}. Storey
et al. investigated the relationship between ``TODO'' comments and
software bugs \cite{storey:08}. Sridhara described a way to detect
up-to-date TODO comments \cite{sridhara:16}. Aman et al. studied
the possibility of commented out codes leading to software bugs
\cite{aman:12, aman:14}.

As for the relationship between code comments and its readers,
Salviulo and Scanniello suggested that novice programmers tend to rely
on comments more than professionals \cite{salviulo:14}. Hirata and
Mizuno examined the relevance of code comments using text filtering
methods \cite{hirata:11}. Some researchers are exploring the idea of
automatic comment generation. Sridhara et al. presented a framework
for automatically generating a Java method description based on
program analysis \cite{sridhara:10}. Wong et al. proposed a way to
generate comments by using programming question sites such as Stack
Overflow \cite{wong:13}.

To our knowledge, code comments are still largely treated
independently from a source code syntax tree. This can be a problem
when a code is automatically refactored by IDE. Sommerlad et al. tried
to address this problem \cite{sommerlad:08}.

\section{Conclusion}

In this paper, we presented our attempt to develop a framework for
collecting and analyzing source code comments in detail.  We proposed
our model of comments, which has three elements: extent, target and
category. We described the definition of each element, and conducted a
manual annotation experiment.  We then presented our attempt to build
classifiers to identify the above three elements using a decision tree
algorithm (C4.5).  The obtained classifiers could recognize these
elements with a reasonable accuracy. We tested our classifiers with
two programming languages (Java and Python).  We applied our
classifiers to various GitHub projects to test our hypothesis that
there is a universal structure in source code comments.

\vspace{1cm}

\bibliographystyle{ACM-Reference-Format}

\end{document}